\documentclass[prl,twocolumn,amsmath,showpacs,superscriptaddress,floatfix]{revtex4}
\usepackage{graphicx}
\begin{document}
\title{Critical Scaling in Linear Response of Frictionless Granular Packings near Jamming}

\author{Wouter G. Ellenbroek} \affiliation{Instituut--Lorentz,
Universiteit Leiden, Postbus 9506, 2300 RA Leiden, The
Netherlands}

\author{Ell\'ak Somfai}
\altaffiliation[Present address: ]{The Rudolf Peierls Centre for Theoretical Physics, 1 Keble Road, Oxford, OX1 3NP, UK}
\affiliation{Instituut--Lorentz, Universiteit Leiden, Postbus 9506, 2300 RA
Leiden, The Netherlands}

\author{Martin van Hecke}
\affiliation{Kamerlingh Onnes Lab, Leiden University, PO box 9504,
2300 RA Leiden, The Netherlands.}

\author{Wim van Saarloos} \affiliation{Instituut--Lorentz,
Universiteit Leiden, Postbus 9506, 2300 RA Leiden, The
Netherlands}

\date{\today}
\begin{abstract}
We study the origin of the scaling behavior in frictionless
granular media above the jamming transition by analyzing their
linear response. The response to local forcing is
non-self-averaging and fluctuates over a length scale that
diverges at the jamming transition. The response to global forcing
becomes increasingly non-affine near the jamming transition. This
is due to the proximity of floppy modes, the influence of which we
characterize by the local linear response. We show that the local
response also governs the anomalous scaling of elastic constants
and contact number.
\end{abstract}
\maketitle
\enlargethispage{0.5cm}
\def\fext{f^\text{ext}}
\newcommand{\z}{z}

The general picture of jamming which was advanced for systems
\cite{jamming,jaeger,epitome,ohern} that form a shear-resistent
solid phase at high densities, is bringing a new perspective to
the deformations of granular and disordered media. A good model
for studying such media are packings of polydisperse weakly
compressible spheres \cite{epitome,ohern}. If we measure pressure
in units of the elastic constants and characteristic radius of the
balls (as we will do below), the relevant limit for granulates is
the small-deformation or, equivalently, the small-pressure limit
in the absence of thermal fluctuations. This limit is also
relevant for weakly compressed emulsions \cite{mason}. We will
focus on the case of frictionless, deformable spherical particles,
and introduce a simple, experimentally accessible and local
measure to characterize the nature of their deformations
\cite{footnote_mu}.

Deformable particles form a stiff jammed phase when the pressure
becomes larger than zero. At the zero pressure jamming point `J',
packings form a ``marginal solid'' and are isostatic, i.e, the
average number of contacts per particle, $\z$, reaches the minimum
$\z^0_{\rm iso}=2d$, needed for a frictionless packing to remain
stable in $d$ dimensions. When the point J is approached by
decreasing the pressure, several surprising scaling relations
emerge: the excess contact number $\Delta \z =\z-\z^0_{\rm iso}$
scales as $\sqrt{\delta}$, with $\delta$ the typical dimensionless
compression of the particles, while the ratio $G/K$ of the shear
modulus $G$ to the compression modulus $K$ scales as $\Delta \z$.
In addition, a diverging time scale $\omega^* \sim \Delta \z$ has
been identified in the density of states of vibrational modes. The
jamming point J thus exhibits features of a critical point
\cite{jaeger,ohern,epitome}.

Since packings at the jamming point are marginal, every additional
broken contact generates a global zero-energy displacement mode, a
so-called \emph{floppy mode}
\cite{epitome,shlomo,wyartlett,wyartpre}. Wyart and coworkers
\cite{wyartlett,wyartpre,wyartthesis} have shown that the scaling
near J is related to those floppy modes, by creating trial modes
for the deformations of weakly jammed solids. These modes are
based on the floppy modes that would occur when along the faces of
cubes of linear size $\ell^*\sim 1 / \Delta \z$  bonds would be
cut.
Even though for jammed systems, truly floppy modes never occur,
their proximity thus governs the scaling just above the jamming
point.

\begin{figure}[bt]
\includegraphics[width=8.5cm]{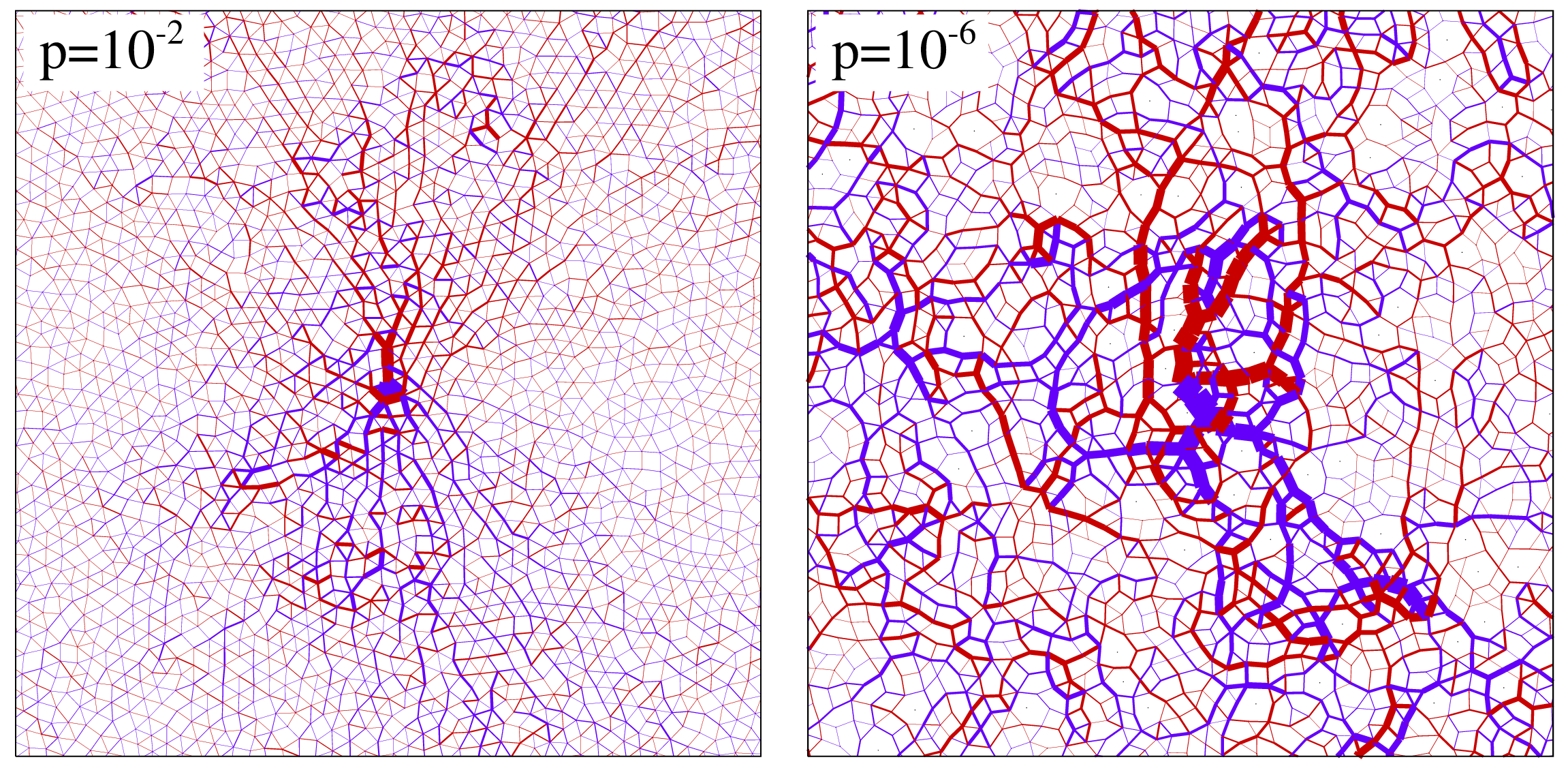}
\caption{(color) Force response networks for a point
loading with pressure as indicated. Blue (red) lines indicate
positive (negative) changes in contact force, the thickness
indicating the amount. The particles themselves are not drawn.}
\label{fig:networks}
\end{figure}

In this Letter we uncover that this proximity of floppy modes
causes an increasingly non-affine response when approaching point
J, and that this response is intimately related to the (anomalous)
scalings of the shear modulus $G$, the excess contact number
$\Delta \z$ and the length scale $\ell^*$. We numerically study
the {\em linear, quasistatic response} of systems near the jamming
transition. The response of granular media has been widely-studied
\cite{serero,gengresp,kasahara,goldnat,glandresp,goldcoarse}, but
not, we believe, systematically as a function of the distance to
the jamming point J. Nor does it seem to have been fully
appreciated that the scaling behavior can essentially be captured
within linear response.

We represent the linear response by relative displacements and
changes in contact forces, and find significant changes with the
distance to point J.
{\em (i)} Fig.~1 illustrates that the response to the loading of a
single grain becomes increasingly disordered over an increasingly
large scale when the jamming transition is approached  --- this
leads to a direct observation of the diverging length scale
$\ell^* \sim 1/ \Delta \z$, shown below. We will show that such a
local force response is not {\em self-averaging}, even though it
is smooth upon {\em ensemble-averaging} and then quantitatively
agrees with continuum elastic behavior. {\em(ii)} The response to
a uniformly applied compression or shear also varies with the
distance to jamming. We introduce the distribution $P(\alpha)$ of
angles $\alpha$ between the bonds and the local deformations as an
indicator of the non-affine nature of the response. Near J,
$P(\alpha)$ becomes strongly peaked around $\alpha = \pi/2$, with
the width and height of the peak scaling with the distance to the
jamming point. Grains then predominantly slide past each other,
which signals an increasingly non-affine response of the material
caused by the proximity of floppy modes, for which
$P(\alpha)=\delta(\alpha-\pi/2)$. {\em(iii)} The component of the
relative displacements perpendicular to the bond vector diverges
upon approaching the jamming point. {\em{(iv)}} Finally, the
$\Delta \z\sim \sqrt{\delta}$ scaling \cite{epitome} is identified
to originate precisely from this increasingly ``sliding''
response.

Hence a simple picture emerges: the influence of floppy modes can
be quantified by the local linear response of the material, which
becomes increasingly non-affine near jamming, in turn  causing
anomalous scaling.

{\em Linear response --- } The response of a jammed granular
medium to external loads has been studied mostly by full scale
molecular dynamics \cite{kasahara,goldnat}. We
calculate it here in linear order from  expansion of energy to
$2^\mathrm{nd}$ order
\begin{equation}\label{Eexpansion}
\Delta E =  \frac{1}{2} \sum_{<ij>}
 k_{ij}\left(
u_{\parallel,ij}^2-\frac{f_{ij}}{k_{ij}r_{ij}}u_{\perp,ij}^2 \!
\right) \! \!.
\end{equation}
Here the sum is over all contacts, $\vec{u}_i$ is the displacement
of particle $i$, and $\vec{u}_{ij}=\vec{u}_j-\vec{u}_i$ the
relative displacement of grains $i$ and $j$, with components
$u_{\parallel,ij}$ and $u_{\perp,ij}$ parallel and perpendicular
to the bond vector $\vec{r}_{ij}=\vec{r}_j-\vec{r}_i$; $k_{ij}$
denotes the stiffness of the contact and $f_{ij}$ its initial
force. The second term proportional to $u_{\perp,ij}^2$ is due to
the transverse motion when the bonds are pre-stressed
($f_{ij}\neq0$). For contact interactions which increase as a
power of the overlap $f_{ij}\sim \delta_{ij}^\beta$, the factor
$f_{ij}/k_{ij}r_{ij}= \delta_{ij}/(\beta r_{ij})$ is of order of
the dimensionless compression $\delta = \delta_{ij}/r_{ij}$, which
is small and which vanishes at the jamming point.

We study 2D packings of $N$ frictionless Hertzian spheres for
which $f_{ij}\sim \delta_{ij}^{3/2}$, where $\delta_{ij}$ is the
overlap between neighboring particles. The confining pressure
ranges from $p=10^{-6}$ to $p=10^{-1}$, in units of the effective
Young modulus of the constituent particles. See
Ref.~\cite{ellakfinally} for details. For each packing, the
expansion~(1) yields the dynamical matrix $M$. Instead of
studying the vibrational dynamics
\cite{wyartlett,wyartpre,ellakjam,tanguy} we obtain here the
quasistatic response to external forces $\fext$ \cite{leonforte}
by solving the linear equation
$ M_{ij,\alpha\beta}u_{j,\beta}=\fext_{i,\alpha}$,
for the $u_j$ and, through the force law, the forces $f_{ij}$.
Here $i,j$ label the particles and $\alpha,\beta$ the  coordinate
axes.

{\em Elastic Moduli} --- We have calculated the elastic moduli
from the linear response by applying an overall compression or
shear and by point-loading a single particle, for packings with
$N=10^3$ and $N=10^4$ respectively. The resulting force fields are
translated into local stress fields \cite{goldcoarse} which are
then ensemble averaged. From fits of the point response
\cite{ellenbroekpowders}, we determine both $K$ and $G$ and
compare these to the values obtained from the response to global
shear and compression. Fig.~2a shows that these two methods agree
very well quantitatively, and that the elastic moduli scale with
pressure as $ K\sim p^{1/3}$, $G\sim p^{2/3}$, in agreement with
earlier results \cite{epitome,makse}.

\begin{figure}[tb]
\includegraphics[width=8.2cm]{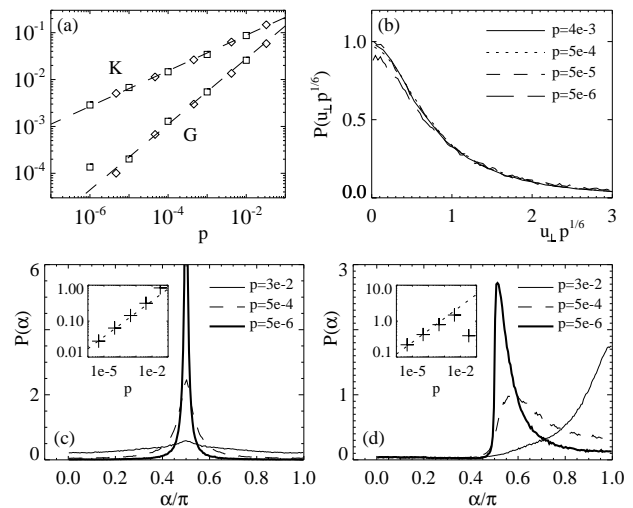}
\caption{(a) Scaling of bulk/shear modulus with pressure, obtained
from point response (squares), and global compression/shear
(diamonds). The fitted exponents are $0.38\pm 0.03$ for the bulk
and $0.70 \pm0.08$ for the shear modulus. (b) Distribution
function of the scaled transverse response for shear (see text).
(c-d) Distribution of the relative displacement angle $\alpha$ for
(c) shear and (d) compression of packings for a range of
pressures. Insets: scaling of the width of the peak $\sim
p^{1/3}$.}
\end{figure}

{\em Non-affinity --- } The typical bond stiffness $k_{ij}$ is
proportional to $p^{1/3}$ for Hertzian contacts. Hence, a simple
estimate for the elastic moduli scaling as $p^{1/3}$ follows under
the affinity assumption that the bond deformations are of order of
the applied deformation. This estimate fails for the shear modulus
$G$, which vanishes faster than $K$ when approaching the jamming
point: $G/K \sim \Delta z$. This has been thought to be caused by
strongly non-affine behavior of the system under shear
\cite{epitome} and the proximity of the floppy modes
\cite{wyartthesis}. We will elucidate now the cause of the scaling
and the influence of the floppy modes via the local deformations
$u_{\parallel,ij}$ and $u_{\perp,ij}$.

As the eigenmodes or snapshots of the response look very disordered
\cite{wyartthesis,ellakfinally,tanguy,leonforte}, it has turned out
to be difficult to find a simple measure to characterize the
non-affinity and the overall floppy mode character. We show now
that proximity of the floppy modes can clearly be identified in
the distribution $P(\alpha)$ of the local angles
$\alpha_{ij}=\mbox{atan} (u_{\parallel,ij}/u_{\perp,ij})$. In a
disordered, isotropic system, one expects $P(\alpha) =
\delta(\alpha-\pi)$ for a purely homogenous compression, and
$P(\alpha) = 1/\pi$ for a purely affine shear. In contrast, for a
floppy mode, $P(\alpha) =  \delta(\alpha -\pi/2)$. This is because
in floppy modes the relative angles between particles change while
the relative distances $r_{ij}$ remain unchanged, as if all the
bonds are replaced by incompressible sticks \cite{shlomo}. Hence,
for a true floppy mode $u_{\parallel,ij}= - u^2_{\perp,ij}/2
r_{ij} \label{fm} + {\mathcal O} (u^4_{\perp,ij} /r_{ij}^3)$
\cite{footnotecancel}.

As the pressure is lowered, $P(\alpha)$ and $P(u_{\perp})$
evidence that the local deformations evolve from near-affine to
extremely non-affine, floppy-mode like (Fig.~2b-d). Indeed, for a
sheared system, $P(\alpha)$ evolves from a flat distribution at
large pressures to a sharply peaked distribution for lower
pressures (Fig.~2c). This peak is located around $\pi/2$ and its
weight approaches 1 --- locally the response becomes more and more
transverse  for $p\to 0$ and $P(\alpha)$ approaches that of a
floppy mode. For a compressed system at large pressures,
$P(\alpha)$ has the ``affine'' peak around $\alpha=\pi$, while for
lower pressures $P(\alpha)$ again develops a sharp peak around
$\alpha = \pi/2$ (Fig.~2d).

Even though the response is far from affine for both compression
and shear, the affine prediction for $K$ holds true while it fails
for $G$. The reason is that for compression, only a finite
fraction of the displacements is essentially transverse and
$P(\alpha)$ remains non-zero away from the peak at $\pi/2$. Since
according to the energy expression (\ref{Eexpansion}) the
compression of bonds given by $u_{\parallel,ij}$ gives the
dominant contribution to $\Delta E$, this is consistent with the
fact that the compression modulus scales with the bond stiffness
$k$: $K\sim k\sim p^{1/3}$. For shear deformations, however, fewer
and fewer bonds contribute to leading order to the energy, and the
weight outside the peak {\em vanishes} as $\Delta z \sim p^{1/3}$,
consistent with the scaling $G/K\sim \Delta z$.

{\em Scaling of $P(\alpha)$ and $P(u_{\perp})$ ---} We can
understand the development of the peak in $P(\alpha)$ from the
balance of terms in the energy expansion (\ref{Eexpansion}).
Focussing on typical values, $ \Delta E \sim k (u_{\parallel}^2 -
\delta u_{\perp}^2)$. Since $k\sim p^{1/3}$ and $\delta \sim
\Delta z^2 \sim p^{2/3}$, balancing the terms we find that
\begin{equation}
\frac{u_{\parallel}}{u_{\perp}} \sim \sqrt{\delta} \sim
p^{1/3}~,\label{parallelperp}
\end{equation}
so that for small $p$, $P(\alpha)$ develops a peak around
$\alpha=\pi/2$, and the width of this peak should scale as
$p^{1/3}$. This is what we find --- see the insets of Figs.~2c,d.

How do the typical values $u_{\parallel}$ and $u_{\perp}$ scale
when we impose a global shear or compression of order $\gamma$ on
the system? Equating the elastic energy densities for compression
and shear, $K \gamma^2$ and $G \gamma^2$, to the energy expansion,
and knowing that the elastic moduli scale as $G \sim p^{2/3}$ and
$K \sim p^{1/3}$, we can predict the scaling of $u_{\parallel}$
and $u_{\perp}$:
\begin{eqnarray}
\mbox{Compression: } & u_{\parallel} \sim p^0 ~~\gamma & \mbox{,
} u_{\perp}
\sim p^{-1/3} \gamma \label{ordercomp}\\
\mbox{Shear: } & u_{\parallel} \sim p^{1/6} \gamma & \mbox{,    }
u_{\perp} \sim p^{-1/6} \gamma\label{ordershear}
\end{eqnarray}
These scaling predictions are well obeyed by our data for small
$p$ --- in Fig.~2b we show this for shear deformations and
$P(u_{\perp})$.

The fact that for fixed $\gamma$ the typical perpendicular
response $u_{\perp}$ {\em diverges} upon approaching the jamming
point, is connected to the disordered nature of the microscopic
response already familiar from the randomly oriented swirl-type
motions \cite{wyartthesis,ellakfinally,tanguy} that characterize
eigenmodes and responses. Of course, in a system of finite size,
$u_{\perp} $ can not diverge. If the cross-over is determined by
the length-scale $\ell^*$ becoming of order the linear system size
$L$, one expects a cross-over scaling $u_{\perp} = L^{1/2}
g(\ell^* /L)$ with $g(w) \sim w^{1/2}$ for $w\ll 1$ and $g\simeq
const$ for $w\gg 1$.
Note also that in the regime $L/\ell^*\ll1$ the response is close
to that of a floppy mode appearing at isostaticity, while our scaling
results apply to the regime $L/\ell^*\gg1$.

{\em $\Delta \z$ scaling} --- The non-affine response also nicely
explains the microscopic origin of the anomalous $\Delta \z \sim
\sqrt \delta$ scaling under compression --- theories assuming
affine deformations give $\Delta \z\sim \delta$. Let us consider a
small compression of the packing with typical bond compression
$u_{\parallel}$. This leads to an infinitesimal change in contact
number
\begin{equation}
\frac{d\Delta \z}{ d\delta } u_{\parallel}\label{deltaZeq}.
\end{equation}
Upon lowering the pressure, the global compression will excite
distorted floppy modes, i.e., for many bonds  $u_{\perp,ij}$ will
be of order $u_{\parallel}/\sqrt{\delta}$ (see Eq.~(\ref{parallelperp})
and Fig.~2d). Moreover, the chance that in such almost
perpendicular displacements one of the particles bumps into a
nearby particle with which it was not in contact yet, will be
proportional to this motion, i.e. to
$u_{\parallel}/\sqrt{\delta}$. Equating this to (\ref{deltaZeq})
then yields $\Delta \z\sim \sqrt{\delta}$. The picture which thus
emerges is that the lower the pressure, the larger the
$u_{\perp,ij}$, and the larger the chances that new contacts are
created.

In earlier papers, it was noted \cite{leogr,epitome,leo06} that the
$\Delta \z\sim \sqrt{\delta}$ scaling with compression was
consistent with a square root divergent term in the correlation
function $g(r)$, if it was assumed that compression would be
essentially affine. As we have seen, however, distortions are not
at all affine near jamming. Our analysis turns this around: it
suggests that the natural coordinates for the floppy mode-like
distortions are the perpendicular displacements, not the radial
ones, and that these generate the square root behavior of $g(r)$
in the radial direction.

\emph{Point response and diverging length scale} --- We finally
return to the point response. Fig.~3a illustrates that the {\em
ensemble} average of such a response conforms to elasticity;
the stress fields fit well and the fitted elastic constants
agree with those obtained from bulk response (Fig.~2a). On the other
hand, individual responses become very disordered and suggest the
occurrence of a large length scale $\ell^*$ when approaching point
J (see Fig.~1).

To extract this length scale, we characterize the response to an
infinitesimal inflation of a single central grain, since the
response to a local {\em directional} force, as shown in Fig.~1
and 3a, is highly anisotropic. We normalize the forces by fitting
the radial stress to the elastic response \cite{neuzel}. We focus
on the radial component of the change in contact force, $df_r$,
calculate $\langle df_r(r)\rangle$ by averaging $df_r$ over
concentric rings, and study the RMS fluctuations
$h(r)\equiv\sqrt{\langle \left[
df_r(r)-\langle{df}_r(r)\rangle\right]^2\rangle}$.

The range over which these fluctuations are felt grows when
approaching the jamming transition (Fig.~3b). When plotted as
function of $r \Delta \z$, the data for $h$ collapses (Fig.~3c).
To our knowledge, this is the first evidence for the existence of
a length scale $\ell^* \sim 1/\Delta \z$ in local response
measurements. Essentially the same characteristic length was shown
by Wyart {\em et al.} to govern the vibrational density of states
\cite{wyartlett,wyartpre}. This scale is identified as the linear
size $\ell^*$ of the largest domain which deforms freely by
pushing on the bonds at its surface. Equating the number of bonds
on the surface ($\sim \ell^{d-1}$) to the number of excess bonds
in the bulk ($\sim \Delta \z\, \ell^d$) yields the maximum size of
such domain $\ell^*\sim 1/ \Delta \z$ \cite{wyartlett,wyartpre}.

Both the average response $\langle{df}_r(r)\rangle$ and the
fluctuations $h(r)$ decay as $1/r^2$  --- the relative
fluctuations do not decay far from the perturbed grain. The
response is {\em not} self-averaging, and there is {\em no} finite
correlation length of fluctuations. The asymptotic value of the
relative fluctuations $h(r)/\langle{df}_r(r)\rangle=h(r)r^2$
grows as $1/\Delta \z^2$
(Fig.~3d) --- one has to coarse grain the response over
increasingly more grains $\mathcal{O}((\Delta \z)^{-4})$ to start
to see convergence to average continuum-like stress response.

\begin{figure}[tb]
\includegraphics[width=8.2cm]{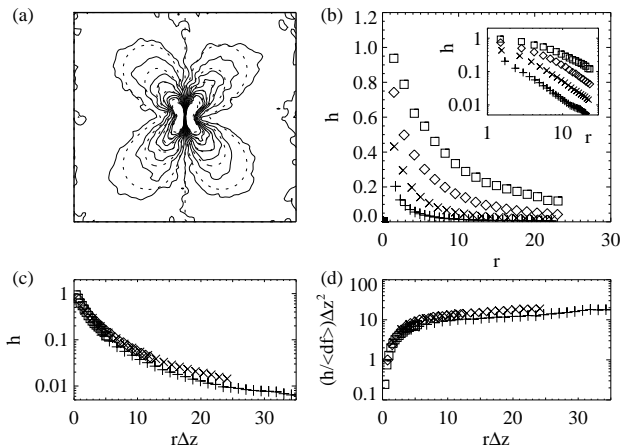}
\caption{(a) Shear stress $\sigma_{xy}$ for point
loading in the center and $p=10^{-3}$. The response is coarse
grained over a grain diameter \cite{goldcoarse} and averaged over
216 realizations (solid) and fitted to linear elasticity (dashed).
(b-d) Spatial decay of the force fluctuations $h$. Squares,
diamonds, crosses and plus signs represent increasing pressures,
ranging from $p=4.6\cdot10^{-5}$ to $p=3.2\cdot10^{-2}$,
respectively. (b) $h$ as function of $r$. (c) Scaling collapse
when $h$  is plotted as function of $r \Delta \z$. (d) Scaling
collapse of relative fluctuations.
}
\label{fig:lengthscale}
\end{figure}

\emph{Outlook} --- Our analysis of the (local) linear response
substantiates and extends the concept that the jammed phase of
weakly compressed frictionless particles is dominated by the
proximity of floppy modes
\cite{shlomo,wyartlett,wyartpre,wyartthesis}. We identified the
increasingly non-affine response to give rise to the scaling
$\Delta \z\sim \sqrt{\delta}$ and presented a direct observation
of the scale $\ell^*\sim 1/\Delta \z$ introduced before
\cite{wyartlett,wyartpre}. The emerging scenario favors a
microscopic, geometric interpretation of these scalings and has
several implications that deserve further study: {\em (i)} What is
the finite size scaling form of $u_{\perp}$? {\em (ii)} What
happens for non-power-law contact interactions such as $f_{ij}\sim
\exp(-\delta_{ij}^{(1-\beta)})$ with $\beta>1$? Our analysis
suggests a scaling $\Delta z \sim \delta^{\beta/2}$. {\em (iii)}
What happens to the square root divergence in $g(r)$
\cite{leogr,epitome,leo06} when the packing algorithm does not allow
for floppy mode-like rearrangements upon annealing, such as may
occur for algorithms based on local rearrangements or packings of
truly hard spheres?

\begin{acknowledgments}
We thank K. Shundyak, J.H. Snoeijer, and M. Depken for helpful
discussions and M. Wyart for critical correspondence which led us
to realize the divergence (\ref{ordercomp},\ref{ordershear}). WGE
acknowledges support from physics foundation FOM and MvH support
from NWO/VIDI.
\end{acknowledgments}


\vspace{-.5cm}
\begin{thebibliography}{99}

\bibitem{jamming} A.J. Liu and S.R. Nagel, Nature \textbf{396}, 21 (1998).

\bibitem{jaeger} H.M. Jaeger,  Physics World \textbf{18} (12), 34 (2005).

\bibitem{epitome} C.S. O'Hern {\em et al.},
Phys.\ Rev.\ E \textbf{68}, 011306 (2003).

\bibitem{ohern} C.S. O'Hern {\em et al.},
Phys.\ Rev.\ Lett.\ \textbf{86}, 111 (2001).

\bibitem{mason} T.G. Mason, J. Bibette and D.A. Weitz,
Phys.\ Rev.\ Lett.\ \textbf{75}, 2051 (1995).

\bibitem{footnote_mu} For frictional packings, the situation is more
complicated: frictional packings are not automatically marginal
solids at jamming, but a critical-like scaling does emerge in the
large-friction limit, see \cite{ellakjam}.

\bibitem{shlomo} S. Alexander, Phys.\ Rep.\ \textbf{296}, 65 (1998).

\bibitem{wyartlett} M. Wyart, S.R. Nagel, and T.A. Witten,
Europhys.\ Lett.\ \textbf{72}, 486-492 (2005).

\bibitem{wyartpre} M. Wyart \emph{et al.},
Phys.\ Rev.\ E \textbf{72}, 051306 (2005).

\bibitem{wyartthesis} M. Wyart, Ann.\ Phys.\ Fr.\ \textbf{30}, 1-96 (2005).

\bibitem{gengresp} J. Geng \emph{et al.},
Phys.\ Rev.\ Lett.\ \textbf{87}, 035506 (2001); Physica D
\textbf{182}, 274 (2003);

\bibitem{serero} D. Serero \emph{et al.}, Eur.\ Phys.\ J. E \textbf{6}, 169 (2001).

\bibitem{kasahara} A. Kasahara and H. Nakanishi,
Phys.\ Rev.\ E \textbf{70}, 051309 (2004).

\bibitem{goldnat} C. Goldenberg and I. Goldhirsch, Nature \textbf{435}, 188 (2005).

\bibitem{glandresp} N. Gland, P. Wang and H.A. Makse, Eur.\ Phys.\ J. E \textbf{20}, 179 (2006).

\bibitem{goldcoarse} I. Goldhirsch and C. Goldenberg, Eur.\ Phys.\ J. E \textbf{9}, 245 (2002).

\bibitem{ellakfinally} E. Somfai \emph{et al.}, Phys.\ Rev.\ E
\textbf{72}, 021301 (2005).

\bibitem{tanguy} A. Tanguy \emph{et al.}, Phys.\ Rev.\ B \textbf{66}, 174205 (2002).

\bibitem{ellakjam} E. Somfai \emph{et al.},
cond-mat/0510506.

\bibitem{leonforte} F. Leonforte \emph{et al.}, Phys.\ Rev.\ B \textbf{70}, 014203 (2004).

\bibitem{ellenbroekpowders} W.G. Ellenbroek {\em et al.},
in \emph{Powders and Grains 2005}, edited by R. Garc\'\i a-Rojo
\emph{et al.} (A.A. Balkema, Rotterdam, 2005), p.~377; and to be
published.

\bibitem{makse} H.A. Makse {\em et al.},
Phys.\ Rev.\ Lett.\ \textbf{83}, 5070 (1999); Phys.\ Rev.\ E
\textbf{70}, 061302 (2004).

\bibitem{footnotecancel} For a floppy mode distortion, the second term in (1)
cancels the linear term $-\sum_{ij}f_{ij}u_{\parallel,ij}$
in the energy expansion. In (1) this term has been
left out on account of force balance in the starting state.

\bibitem{leogr} L.E. Silbert \emph{et al.}, Phys.\ Rev.\ E
\textbf{65}, 031304 (2002).

\bibitem{leo06} L.E. Silbert, A.J. Liu and S.R. Nagel,
Phys.\ Rev.\ E \textbf{73}, 041304 (2006).

\bibitem{neuzel} A
very small fraction (typically 1\%) of the responses is so
disordered that the overlap with continuum behavior is minute
(often because a cluster of particles far away shifts
dramatically) --- these have been omitted.

\end{thebibliography}
\end{document}